\documentclass[twocolumn,aps,preprintnumbers,prl,10pt,superscriptaddress]{revtex4-1}
\usepackage{graphicx}
\usepackage{amsmath}
\usepackage{multirow}

\begin{document}

\preprint{ZU-TH 09/11}

\title{Search for anomalous top quark production at the early LHC}

\author{Jun Gao}
\author{Chong Sheng Li}\email{csli@pku.edu.cn}
\affiliation{Department of Physics and State Key Laboratory of Nuclear Physics and
  Technology, Peking University, Beijing 100871, China}
\author{Li Lin Yang}
\affiliation{Institute for Theoretical Physics, University of Z\"urich, CH-8057 Z\"urich,
  Switzerland}
\author{Hao Zhang}
\affiliation{Department of Physics and State Key Laboratory of Nuclear Physics and
  Technology, Peking University, Beijing 100871, China}


\begin{abstract}
  We present a detailed study of the anomalous top quark production with subsequent decay
  at the LHC induced by model-independent flavor-changing neutral-current couplings,
  incorporating the complete next-to-leading order QCD effects. Our results show that,
  taking into account the current limits from the Tevatron, the LHC with $\sqrt{s}=7$~TeV
  may discover the anomalous coupling at 5$\sigma$ level for a very low integrated
  luminosity of $61$~pb$^{-1}$. The discovery potentials for the anomalous couplings at
  the LHC are examined in detail. We also discuss the possibility of using the charge
  ratio to distinguish the $tug$ and $tcg$ couplings.
\end{abstract}

\maketitle

The CERN Large Hadron Collider (LHC) is currently operating at a center-of-mass (c.m.)
energy of 7\,TeV. Even with a relatively low integrated luminosity ($\sim\,40\,{\rm
pb}^{-1}$), it has already delivered many interesting results, including new tests on the
standard model (SM) in both electroweak and strong interacting sectors, as well as
constraints on new physics models beyond the SM. In particular, both the ATLAS and the
CMS collaborations have measured the cross section for top quark pair production with a
precision of around 10\%. With more data being collected, it can be expected that the top
quark properties will be well measured in the near future, and exotic physics in the top
quark sector can also be potentially probed.

In the SM, flavor-changing neutral-currents (FCNC) in the top quark sector are strongly
suppressed. However, many new physics models can induce large FCNC couplings of the top
quark with a light up-type quark and a gluon, which can be possibly detected at the
LHC~\cite{Tait:2000sh, AguilarSaavedra:2004wm}. Such couplings can be incorporated into
the model-independent effective Lagrangian \cite{Malkawi:1995dm, Hosch:1997gz}
\begin{align*}
  \mathcal{L} = g_s \sum_{q=u,c} \frac{\kappa_{tqg}}{\Lambda} \bar{t} \sigma^{\mu\nu} T^a
  (f_q^LP_L+f_q^RP_R) q G^a_{\mu\nu} + h.c. \, ,
\end{align*}
where $\kappa_{tqg}/\Lambda$ are real numbers representing the strength of the couplings,
and $f_{q}^{L,R}$ are chiral parameters normalized to $|f_{q}^{L}|^2+|f_{q}^{R}|^2=2$.
Both the CDF and D0 collaborations at the Tevatron have searched for processes induced by
these operators and provided constraints on the anomalous couplings \cite{Aaltonen:2008qr,
  Abazov:2010qk}. The most stringent one-dimensional exclusion limit (assuming only one
coupling is non-zero) is given by \cite{Abazov:2010qk}
\begin{align}
  \kappa_{tug}/\Lambda < 0.013\,{\rm TeV}^{-1} \, , \quad \kappa_{tcg}/\Lambda <
  0.057\,{\rm TeV}^{-1} \, , \label{eq:limit}
\end{align}
at the 95\% confidence level (C.L.). The above anomalous couplings can induce various rare
processes at hadron colliders. Among them, the most interesting one is direct top quark
production, where a single top quark is produced without any additional particle. The
signature of this process is different from the single top production in the SM (where the
top quark is always accompanied by other particles). Given the couplings allowed by the
Tevatron search, the production rate for this process can still be large at the LHC, which
makes it a promising channel to search for new physics in the flavor sector. Any
observation of this characteristic process definitely indicates the existence of the $tqg$
anomalous couplings, and the underlying new physics.

There have been several analyses in the literature of direct top quark production at the
LHC at the leading order (LO) \cite{Hosch:1997gz, AguilarSaavedra:2004wm,
  Ferreira:2005dr}. The next-to-leading order (NLO) QCD correction to the total cross
section of this process has also been calculated in \cite{Liu:2005dp}. However, there is
no detailed phenomenological study based on the NLO result. Also, the previous LO studies
\cite{Hosch:1997gz, AguilarSaavedra:2004wm} only focused on the LHC with
$\sqrt{s}=14$\,TeV. Moreover, they did not include the SM single top quark production in
the background processes, and therefore underestimated the background rate. The SM single
top quark production can mimic the signal process if the additional jet is not
reconstructed. At the NLO, the signal process can also emit an additional jet which makes
the single top background more prominent. With these considerations, it is therefore very
important to perform an analysis taking into account the NLO QCD effects and all the SM
backgrounds for the early LHC search of the anomalous couplings with $\sqrt{s}=7$\,TeV.
Besides, in order to provide a more complete NLO prediction, the QCD effects in the top
quark decay process should also be included. While the QCD correction to the decay
process does not alter the inclusive rate, it may change the signal acceptance
significantly when kinematic cuts are applied. In this Letter, we present a detailed
study of the direct top quark production with subsequent decay at the LHC, including NLO
QCD corrections for both the production part and decay part. The SM backgrounds and the
LHC discovery potential of the anomalous couplings are also examined in detail.

The NLO QCD corrections to the direct top quark production with subsequent decay can be
factorized into two independent gauge invariant parts, i.e., the top quark production at
NLO with subsequent decay at LO, and production at LO with subsequent decay at NLO, by
using the modified narrow width approximation incorporating the finite width effects as
in the previous studies of the SM single top quark
production~\cite{Cao:2004ap,Campbell:2004ch}. Note that if we only consider the inclusive
rate of the process, the NLO QCD corrections to the decay part will have no influence
since the branching ratio of the top quark decay into $W$ boson is always 100\% at both
the LO and the NLO (the branching ratios of the rare decays $t \to
gq$~\cite{Zhang:2008yn} are negligible here considering the Tevatron limits). But for the
production rate after applying final state kinematic cuts, the NLO QCD corrections to the
decay part can have significant contributions as we will show below.

In the following analysis we only consider the leptonic decay (with two lepton flavors) of
the top quark since the hadronic decay mode suffers from large QCD backgrounds. And also
we need to use $b$-tagging to suppress the large $W$+jet backgrounds. We use the following
basic selection cuts
\begin{gather}
  p_{Tl} > 20\,{\rm GeV} \, , \quad p_{Tb} > 50\,{\rm GeV} \, , \quad
  \rlap{\hspace{0.08em}/}{E}_{T} > 30\,{\rm GeV} \, , \nonumber
  \\
  |\eta_l| < 2.4 \, , \quad |\eta_b| < 2.0 \, , \quad \Delta R_{bl} > 0.7 \,
  , \label{dtopcut1}
\end{gather}
and the anti-$k_T$ jet algorithm with the jet radius parameter $D=0.7$. Here
$\rlap{\hspace{0.08em}/}{E}_{T}$ is the missing transverse energy, $p_{Tl(b)}$ and
$\eta_{l(b)}$ are the transverse momentum and pseudorapidity of the lepton and $b$-jet,
respectively, and $\Delta R=\sqrt{(\Delta\eta)^2+(\Delta\phi)^2}$ measures the angular
separation between the two objects, with $\phi$ being the azimuthal angle. We also demand
$\Delta R_{jl}>0.7$ if there is an additional jet with $p_T>20$\,GeV and $|\eta|<3$.
To further reduce the $W$+jet backgrounds we require the mass $m_{\text{r,top}}$ and
rapidity $y_{\text{r,top}}$ of the reconstructed top quark to satisfy
\begin{align}
  160\,{\rm GeV} < m_{\text{r,top}} < 185\,{\rm GeV} \, , \quad |y_{\text{r,top}}|>0.3 \,
  . \label{dtopcut2}
\end{align}
With all the above selection conditions, we can get a stable $b$-tagging efficiency of
50\% while the misidentification rates for other jets are, 8\% for charm quark, 0.2\% for
gluon and other light quarks~\cite{Aad:2009wy}. Another main background of the signal
process is the SM $t$-channel single top quark production, where the associated jet
prefers to be in the forward region. In order to suppress its contribution we will use a
jet veto just as in the case of SM Higgs boson search in the $WW$
channel~\cite{Anastasiou:2007mz}. Here we require the final state contains exact one
charged lepton and one $b$-jet, and events containing additional jets with $p_T>30$\,GeV
and $|\eta|<3$ will be vetoed. Moreover, to account for the resolution of the detectors,
we use the following smearing parameters for lepton, jets and missing transverse
momentum~\cite{Aad:2009wy}
\begin{align}
  &\Delta E_l/E_l = 0.1/\sqrt{E_l/{\rm GeV}} \oplus 0.007 \, , \nonumber
  \\
  &\Delta \eta_l = \Delta \phi_l = 0.001 \, , \nonumber
  \\
  &\Delta E_{b(j)}/E_{b(j)} = 0.5/\sqrt{E_{b(j)}/{\rm GeV}} \oplus 0.02 \, , \nonumber
  \\
  &\Delta \eta_{b(j)} = \Delta \phi_{b(j)} = 0.01 \, ,\nonumber
  \\
  &\Delta\rlap{\hspace{0.08em}/}{E}_{T,x(y)} = 0.46\sqrt{H_T/{\rm GeV}} \, ,
\end{align}
where $E_{l,b,j}$, $\eta_{l,b,j}$ and $\phi_{l,b,j}$ are the energy, pseudorapidity and
azimuthal angle of the lepton, $b$-jet and the additional jet, respectively; $H_T$ is the
scalar sum of the transverse energy of all the lepton and jets. The symbol $\oplus$
denotes adding in quadrature.

We now present the numerical results of our study. First, we give the total cross
sections for the signal process of the direct top quark production. The CTEQ6L1
(CTEQ6.6)~\cite{Nadolsky:2008zw} parton distribution function (PDF) and the corresponding
running QCD coupling constant are used in the LO (NLO) calculations. We choose the
anomalous couplings $\kappa_{tug}/\Lambda=\kappa_{tcg}/\Lambda=0.01\,{\rm TeV}^{-1}$, and
the renormalization and factorization scales to be the top quark mass $m_t=173.1$\,GeV,
unless otherwise specified. We use L and R to denote the pure left-handed couplings
($f^R_g=0$) and pure right-handed couplings ($f^L_g=0$), respectively. Table~\ref{dtopt1}
shows the inclusive cross sections for the top (anti-top) quark production with
subsequent leptonic decay, without applying any kinematic cuts. The scale uncertainties
shown are calculated by varying the renormalization and factorization scales between
$m_t/2$ and $2m_t$. It can be seen that the NLO corrections reduce the scale
uncertainties from about 10\% to about 7\%. We have also estimated the PDF uncertainties
by using the 44 error sets of the CTEQ6.6 PDFs, and the values are within $\pm 4\%$ for
the NLO inclusive cross sections.
\begin{table}[h!]
  \begin{center}
      \begin{tabular}{c|cc}
        \hline \hline
        & $\sigma_{\text{LO}}$ [pb] & $\sigma_{\text{NLO}}$ [pb]
        \\ \hline
        $tug$ (L/R) & $1.12^{+0.10}_{-0.09}$ ($0.202^{+0.020}_{-0.018}$) &
        $1.48^{+0.10}_{-0.09}$ ($0.285^{+0.018}_{-0.017}$)
        \\ \hline
        $tcg$ (L/R) & $0.105^{+0.011}_{-0.011}$ (---) & $0.159^{+0.012}_{-0.012}$ (---)
        \\ \hline \hline
      \end{tabular}
  \end{center}
  \vspace{-3ex}
  \caption{\label{dtopt1}The total cross sections for the direct top quark production with
    subsequent leptonic decay at the LHC ($\sqrt{s}=7$\,TeV) without applying any
    kinematic cuts. The numbers in brackets correspond to anti-top quark production.}
\end{table}

\begin{table}[h!]
  \begin{center}
    \begin{tabular}{cc|cccc}
      \hline \hline
      & & $\sigma_{\text{LO}}$ [fb] & $K_{\text{pro}}$ & $K_{\text{tot}}$ & $K_{\text{veto}}$
      \\ \hline
      \multirow{2}{*}{$tug$} & (L) & 148 (40.5) & 1.36 (1.32) &1.27 (1.23) & 0.91 (0.96)
      \\
      & (R) & 164 (41.3) & 1.34 (1.32) &1.24 (1.22) & 0.90 (0.94)
      \\ \hline
      \multirow{2}{*}{$tcg$} & (L) & 21.7 (---) & 1.37 (---) & 1.27 (---) & 0.96 (---)
      \\
      & (R) & 21.8 (---) & 1.36 (---) & 1.26 (---) & 0.95 (---)
      \\ \hline \hline
    \end{tabular}
  \end{center}
  \vspace{-3ex}
  \caption{\label{dtopt2}The LO cross sections and $K$-factors of the direct top quark
    production at the LHC ($\sqrt{s}=7$\,TeV) with the kinematic cuts Eqs.~(\ref{dtopcut1}) and
    (\ref{dtopcut2}) applied, and jet veto in addition for $K_{\text{veto}}$.}
\end{table}
When constraints on the final state kinematics are imposed, we need to include the NLO QCD
corrections to the decay process as well. We find that these corrections will reduce the
cross sections by 10\%. This can be understood since the emission of an extra gluon can
broaden the reconstructed top quark mass distribution, and can also make the $b$-jet
softer. The corresponding results are shown in Table~\ref{dtopt2}, where the $K$-factors
are defined as the ratio of the NLO cross sections to the LO ones. Here the kinematics
cuts Eqs.~(\ref{dtopcut1}) and (\ref{dtopcut2}) are always imposed. $K_{\text{pro}}$
includes only the NLO corrections to the production part , $K_{\text{tot}}$ includes also
the corrections to the decay part, and $K_{\text{veto}}$ has in addition jet veto applied.
The differences between the results of the left-handed and the right-handed couplings are
due to the spin effects of the top quark. After applying the jet veto, the complete NLO
QCD corrections are further reduced to $-4\% \sim -10\%$ as shown in Table~\ref{dtopt2}
since most of the contributions from the real corrections are dropped. And the scale
uncertainties of the NLO cross sections are within $\pm 4\%$ of the center values. By
comparing the results in Table~\ref{dtopt1} and \ref{dtopt2}, we can see that the
differences of the $K$-factors between the inclusive rate and the one with the kinematic
cuts and jet veto can be over 40\%. Finally, after taking into account all the cuts and
the $b$-tagging efficiency, the NLO cross sections of the signal process can be written as
\begin{align}
  \sigma_{\text{signal}} = \sum_{q=u,c} \left( \frac{\kappa_{tqg}}{\Lambda} \right)^2
  (b_{qL}|f_q^L|^2+b_{qR}|f_q^R|^2) \, ,
\end{align}
with the coefficients $b_{qL(R)}$ given in Table~\ref{dtopt3}.
\begin{table}[h!]
  \begin{center}
    \begin{tabular}{cc|cccc}
      \hline \hline
      \multicolumn{2}{c|}{[pb$\cdot{\rm TeV}^2$]} & $b_{uL}$ & $b_{uR}$ & $b_{cL}$ & $b_{cR}$
      \\ \hline
      \multirow{2}{*}{7\,TeV} & $l^+$ & 338 & 368 & \multirow{2}{*}{52.5} & \multirow{2}{*}{51.8}
      \\
      & $l^-$ & 97.0 & 96.8 & &
      \\ \hline
      \multirow{2}{*}{14\,TeV} & $l^+$ & 588 & 625 & \multirow{2}{*}{167} & \multirow{2}{*}{164}
      \\
      & $l^-$ & 272 & 272 & &
      \\ \hline \hline
    \end{tabular}
  \end{center}
  \vspace{-3ex}
  \caption{\label{dtopt3}Values of the coefficients $b_{qL(R)}$.}
\end{table}

We now turn to the SM backgrounds, which are calculated with a modified version of the
MCFM~\cite{Campbell:2004ch} package and crosschecked with MadGraph~\cite{Alwall:2007st}.
The PDF choice here is the same as in the signal process while the scale is set to $m_Z$
and $m_t$ for the $W$ boson and top quark related processes, respectively. The main
contributions arise from the SM $t$-channel single top quark production, the $W$+jet
production and $W+bb(bq)$ production as shown in Table~\ref{dtopt4}, where for the sum in
the last column the $b$-tagging efficiency and misidentification rates have been taken
into account. The $t$-channel single top and $W+c$ processes are calculated up to the NLO
in QCD. Other contributions from $W+cc(cq)$ production, the top quark pair production,
the vector boson pair production and other single top quark production channels can all
be neglected according to our calculation.
\begin{table}[h!]
  \begin{center}
    \begin{tabular}{cc|ccccc}
      \hline \hline
      \multicolumn{2}{c|}{[pb]} & $Wu(d,s,g)$ & $Wc$ & $Wbb(bq)$ &single top  & total
      \\ \hline
      \multirow{2}{*}{7 TeV} & $l^+$ & 33 & 2.67 & 0.14 & 0.632 & 0.666
      \\
      & $l^-$ & 18 & 2.81 & 0.077 & 0.320 & 0.460
      \\ \hline
      \multirow{2}{*}{14 TeV} & $l^+$ & 69 & 7.60 & 0.28 & 2.27 & 2.02
      \\
      & $l^-$ & 47 & 8.16 & 0.18 & 1.33 & 1.50
      \\ \hline \hline
    \end{tabular}
  \end{center}
  \vspace{-3ex}
  \caption{\label{dtopt4}The cross sections for the SM background processes.}
\end{table}

\begin{figure}[h!]
  \begin{center}
    \includegraphics[width=0.3\textwidth]{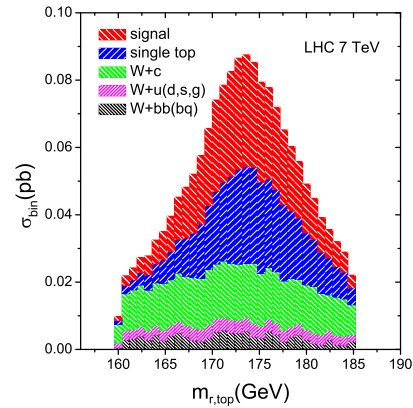}
  \end{center}
  \vspace{-3ex}
  \caption{\label{dtopf1} The reconstructed top quark mass distribution from the signal
    and background processes, assuming $f^L_q=0$ and $\kappa_{tcg}/\Lambda=0.05\,{\rm TeV}^{-1}$,
    $\kappa_{tug}/\Lambda=0$.}
  \vspace{-3ex}
\end{figure}

In Fig.~\ref{dtopf1} we show the reconstructed top quark mass distribution from the
signal and background processes. Here and below we have added the positive and negative
charged lepton contributions together. It can be seen that the SM $t$-channel single top
quark production contributes over 40\% of the total backgrounds and is hard to be further
suppressed since it has almost the same kinematics as the signal process (both from top
quark decay).

\begin{figure}[t!]
  \begin{center}
    \includegraphics[width=0.3\textwidth]{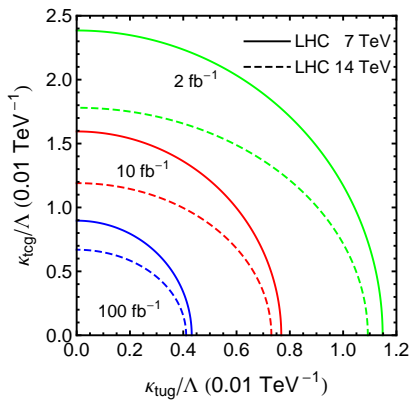}
  \end{center}
  \vspace{-3ex}
  \caption{\label{dtopf2} The $5\sigma$ discovery limits of the anomalous couplings for
    the LHC with different c.m. energies and luminosities.}
\end{figure}

To further study the LHC discovery potential of the anomalous couplings, we choose
$f^L_q=f^R_q=1$ as in the previous experimental analysis at the Tevatron. If we take the
anomalous couplings equal to the one-dimensional upper limits in Eq.~(\ref{eq:limit}), we
find that, for the LHC with $\sqrt{s}=7$\,TeV, the needed integrated luminosity for a
$5\sigma$ discovery ($N_S/\sqrt {N_B}=5$) of the signal process is $61\,{\rm pb}^{-1}$
($1.22\,{\rm fb}^{-1}$) for the $tcg$ ($tug$) coupling, which means that we may expect
the observation of this rare process at the early stage of the LHC. Moreover, in
Fig.~\ref{dtopf2} we show the $5\sigma$ discovery limits of the anomalous couplings for
the LHC with different integrated luminosities and c.m. energies. We find that with a
luminosity of $2\,(10)\,{\rm fb}^{-1}$ and $\sqrt{s}=7$\,TeV, the LHC can detect the
anomalous couplings down to $0.0115\,(0.0077)\,{\rm TeV}^{-1}$ and $0.024\,(0.016)\,{\rm
TeV}^{-1}$ for the $tug$ and $tcg$ couplings, respectively, assuming only one coupling is
non-zero. The increase of the c.m. energy will raise the discovery potential for the
$tcg$ coupling significantly. Assuming that the LHC could reach its designed energy of
$14$\,TeV and collect an integrated luminosity of $100\,{\rm fb}^{-1}$, we can see that
the one-dimensional discovery limits of the anomalous couplings can be as low as
$0.0041\,{\rm TeV}^{-1}$ and $0.0067\,{\rm TeV}^{-1}$. The LHC shows a great improvement
for the detection of the $tcg$ anomalous coupling as compared to the Tevatron due to the
large enhancement of the charm quark PDF.
\begin{figure}[h!]
  \begin{center}
    \includegraphics[width=0.22\textwidth]{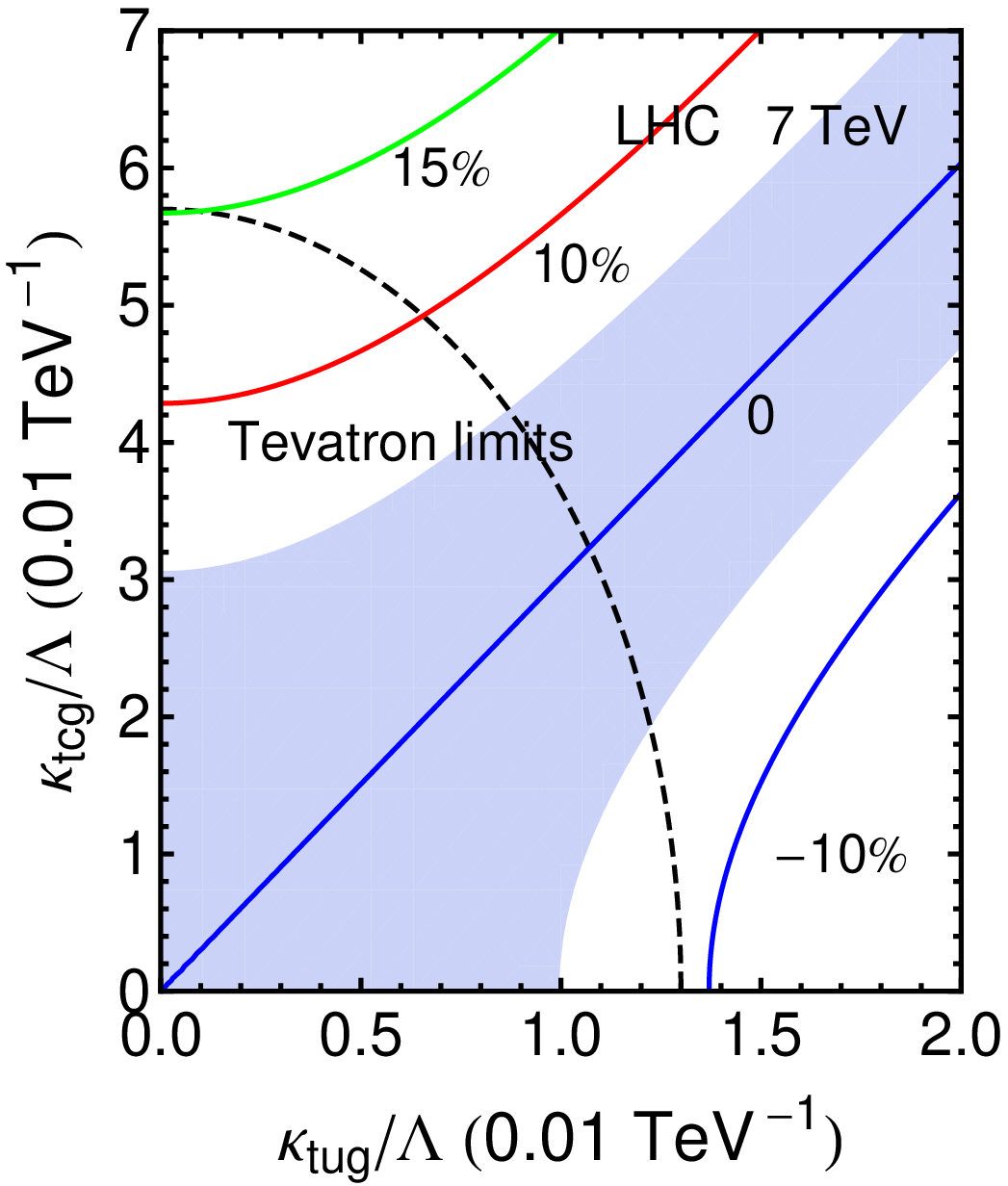}
    \includegraphics[width=0.22\textwidth]{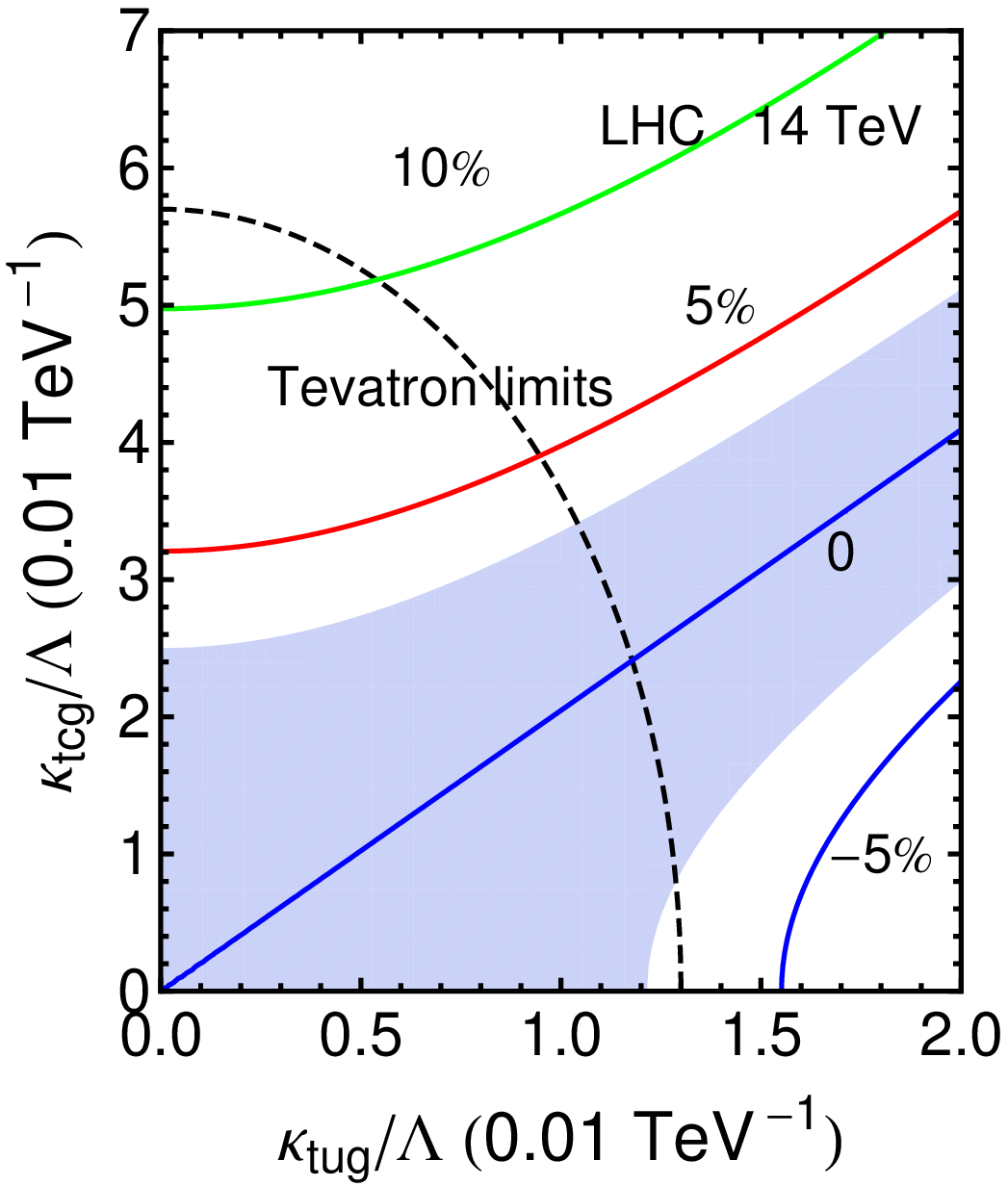}
  \end{center}
  \vspace{-3ex}
  \caption{\label{dtopf6} Contour plot for the relative deviation of the ratio
    $\sigma(l^-)/\sigma(l^+)$ induced by the anomalous couplings, the shadow regions
    indicate the $\pm 3\sigma$ statistic fluctuation of the SM background values assuming
    an integrated luminosity of $10\,{\rm fb}^{-1}$.}
  \vspace{-3ex}
\end{figure}

Once the direct top quark production is observed, it is important to determine whether it
comes from the up quark or charm quark initiated process for the understanding of the
underlying new physics. In Fig.~\ref{dtopf6} we plot the relative deviation of the charge
ratio $\sigma(l^-)/\sigma(l^+)$ induced by the anomalous couplings as compared to the SM
backgrounds. The shadowed regions indicate the $\pm 3\sigma$ statistic fluctuation of the
SM background values assuming an integrated luminosity of $10\,{\rm fb}^{-1}$ and the
dashed lines represent the current two-dimensional exclusion limits from
Tevatron~\cite{Abazov:2010qk}. Since the $tcg$ coupling contributes equally to the top
and anti-top quark production, it prefers a larger value of the ratio as compared to the
SM backgrounds. When the LHC collects enough data, a precision measurement of the charge
ratio can be expected, which can be used to distinguish the $tug$ and $tcg$ couplings
together with the total cross section measurements. Furthermore, we also emphasize that
if the anomalous couplings are pure left(right)-handed, the top quark spin direction is
parrel (opposite) to the incident quark direction, which can be determined through the
boost direction of the reconstructed top quark only for the $tug$ coupling. Thus it may
be possible to determine the chiral structure of the $tug$ anomalous coupling by studying
the angular distribution of the charged lepton, which will be discussed in detail
elsewhere.

In conclusion, we have performed a detailed study of direct top quark production with
subsequent decay including the complete NLO QCD effects at the LHC. We show that after
using the kinematic cuts and jet veto in the final states, the QCD corrections of the
signal process become negative while the ones for the inclusive rate are large and
positive, which may have significant influence to the precision measurement of the
anomalous couplings at the LHC. Assuming anomalous couplings equal to the current
experimental limits, we find for the LHC with $\sqrt{s}=7$~TeV that the integrated
luminosity needed for a $5\sigma$ discovery of the anomalous top quark production is as
low as $61\,{\rm pb}^{-1}$, and can be easily accomplished at the early stage of the LHC.
Besides, with an integrated luminosity of $2(10)\,{\rm fb}^{-1}$ and $\sqrt s=7\,{\rm
  TeV}$ the LHC can probe the anomalous couplings down to $0.0115(0.0077)\,{\rm TeV}^{-1}$
and $0.024(0.016)\,{\rm TeV}^{-1}$ for the $tug$ and $tcg$ couplings, respectively,
assuming only one coupling is non-zero.

\begin{acknowledgments}
  This work was supported in part by the National Natural Science Foundation of China,
  under Grants No.11021092 and No.10975004, by the European Commission through the
  `LHCPhenoNet' Initial Training Network PITN-GA-2010-264564, and by the Schweizer
  Nationalfonds under grant 200020-124773.
\end{acknowledgments}

\end{document}